\documentclass{natureprintstyle}

\usepackage{graphicx}
\usepackage{amssymb}
\usepackage{epstopdf}

\usepackage{apjfonts}
\usepackage{amssymb}
\usepackage{graphicx}
\usepackage{ulem}
\usepackage[utf8]{inputenc}
\usepackage[T1]{fontenc}
\usepackage{url}
\usepackage{hyperref}
\usepackage{mdwlist}
\usepackage{amsmath} 
\usepackage{xfrac}
\usepackage{enumitem}
\usepackage{bm}
\makeatletter
\def\url@leostyle{%
 \@ifundefined{selectfont}{\def\UrlFont{\sf}}{\def\UrlFont{\small\ttfamily}}}
\makeatother
\newcommand{\ls}{{_<\atop^{\sim}}}
\newcommand{\gs}{{_>\atop^{\sim}}}
\def \spose#1{\hbox  to 0pt{#1\hss}}  
\def \ls{\mathrel{\spose{\lower 3pt\hbox{$\sim$}}\raise  2.0pt\hbox{$<$}}}
\def \gs{\mathrel{\spose{\lower  3pt\hbox{$\sim$}}\raise 2.0pt\hbox{$>$}}}
\newcommand{\Ha}{\hbox{{\rm H}$\alpha$}}

\newcommand{\OIII}{\hbox{[{\rm O}\kern 0.1em{\sc iii}]}}
\newcommand{\OII}{\hbox{[{\rm O}\kern 0.1em{\sc ii}]}}
\newcommand{\NII}{\hbox{[{\rm N}\kern 0.1em{\sc ii}]}}
\newcommand{\SII}{\hbox{[{\rm S}\kern 0.1em{\sc ii}]}}
\newcommand{\HII}{\hbox{{\rm H}\kern 0.1em{\sc ii}}}
\newcommand{\angstrom}{\textup{\AA}}

\newcommand{\msun}{M_{\odot}}

\title{Suppressing star formation in quiescent galaxies with supermassive black hole winds}

\author{Edmond Cheung$^{1}$ , Kevin Bundy$^{1}$,
Michele Cappellari$^{2}$,
S\'{e}bastien Peirani$^{1,3}$, 
Wiphu Rujopakarn$^{1,4}$, 
Kyle Westfall$^{5}$, 
Renbin Yan$^{6}$, 
Matthew Bershady$^{7}$, 
Jenny E. Greene$^{8}$, 
Timothy M. Heckman$^{9}$, 
Niv Drory$^{10}$, 
David R. Law$^{11}$,
Karen L. Masters$^{5}$,
Daniel Thomas$^{5}$,
David A. Wake$^{7,12}$,
Anne-Marie Weijmans$^{13}$,
Kate Rubin$^{14}$, 
Francesco Belfiore$^{15,16}$,
Benedetta Vulcani$^{1}$, 
Yan-mei Chen$^{17}$, 
Kai Zhang$^{6}$, 
Joseph D. Gelfand$^{18,19}$,
Dmitry Bizyaev$^{20,21}$,
A. Roman-Lopes$^{22}$,
\& Donald P. Schneider$^{23,24}$ 
}

\begin{document}

\maketitle


\begin{abstract}
Quiescent galaxies with little or no ongoing star formation dominate the galaxy population above $M_{*}\sim 2 \times 10^{10}~M_{\odot}$, where their numbers have increased by a factor of $\sim25$ since $z\sim2$\cite{bell04,bundy06,faber07,ilbert10}. Once star formation is initially shut down, perhaps during the quasar phase of rapid accretion onto a supermassive black hole\cite{dimatteo05,hopkins06,heckman14}, an unknown mechanism must remove or heat subsequently accreted gas from stellar mass loss\cite{ciotti97} or mergers that would otherwise cool to form stars\cite{benson03,booth13}. Energy output from a black hole accreting at a low rate has been proposed\cite{croton06,bower06,ciotti10}, but observational evidence for this in the form of expanding hot gas shells is indirect and limited to radio galaxies at the centers of clusters\cite{fabian06,fabian12}, which are too rare to explain the vast majority of the quiescent population\cite{lin07}. Here we report bisymmetric emission features co-aligned with strong ionized gas velocity gradients from which we infer the presence of centrally-driven winds in typical quiescent galaxies that host low-luminosity active nuclei.  These galaxies are surprisingly common, accounting for as much as $10\%$ of the population at $M_* \sim 2 \times 10^{10}~ M_{\odot}$. In a prototypical example, we calculate that the energy input from the galaxy's low-level active nucleus is capable of driving the observed wind, which contains sufficient mechanical energy to heat ambient, cooler gas (also detected) and thereby suppress star formation.
\end{abstract}


Using optical imaging spectroscopy from the Sloan Digital Sky Survey-IV Mapping Nearby Galaxies at Apache Point Observatory\cite{bundy15} (SDSS-IV MaNGA) program, we define a new class of quiescent galaxies (required to have red rest-frame colors, $NUV-r>5$) that is characterized by the presence of narrow bisymmetric patterns in equivalent width (EW) maps of strong emission lines, such as H$\alpha$ and $\OIII$.  Our selection employs multiband imaging to exclude galaxies with dust lanes and other disk signatures. The observed enhanced emission features are oriented randomly with respect to the optical surface brightness morphology, but roughly align with strong, systematic velocity gradients as traced by the ionized gas emission lines. The gas velocity fields in these galaxies are decoupled from their stellar motions. These galaxies are surprisingly common among the quiescent population, accounting for $\sim$10\% of quiescent galaxies with $\log~M_*/M_{\odot} \sim 10.3$.

\let\thefootnote\relax\footnote{
\begin{affiliations}
$^{1}$Kavli Institute for the Physics and Mathematics of the Universe (WPI), The University of Tokyo Institutes for Advanced Study, The University of Tokyo, Kashiwa, Chiba 277-8583, Japan \\
$^{2}${Sub-department of Astrophysics, Department of Physics, University of Oxford, Denys Wilkinson Building, Keble Road, Oxford OX1 3RH, UK}\\
$^{3}${Institut d'Astrophysique de Paris (UMR 7095: CNRS and UPMC), 98 bis Bd Arago F-75014 Paris, France} \\
$^{4}${Department of Physics, Faculty of Science, Chulalongkorn University, 254 Phayathai Road, Pathumwan, Bangkok 10330, Thailand}\\
$^{5}${Institute for Cosmology and Gravitation, University of Portsmouth, Dennis Sciama Building, Burnaby Road, Portsmouth PO1 3FX}\\
$^{6}${Department of Physics and Astronomy, University of Kentucky, 505 Rose Street, Lexington, KY 40506-0055, USA}\\
$^{7}${Department of Astronomy, University of Wisconsin-Madison, 475 North Charter Street, Madison, WI 53706, USA}\\
$^{8}${Department of Astrophysical Sciences, Princeton University, Princeton, NJ 08544, USA}\\
$^{9}${Center for Astrophysical Sciences, Department of Physics \& Astronomy, The Johns Hopkins University, Baltimore, Maryland 21218}\\
$^{10}${McDonald Observatory, Department of Astronomy, University of Texas at Austin, 1 University Station, Austin, TX 78712-0259, USA}\\
$^{11}${Space Telescope Science Institute, 3700 San Martin Drive, Baltimore, MD 21218, USA}\\
$^{12}${Department of Physical Sciences, The Open University, Milton Keynes, MK7 6AA, UK}\\
$^{13}${School of Physics and Astronomy, University of St Andrews, North Haugh, St Andrews, Fife KY16 9SS, UK}\\
$^{14}${Harvard-Smithsonian Center for Astrophysics, 60 Garden Street, Cambridge, MA 02138, USA}\\
$^{15}${Cavendish Laboratory, University of Cambridge, 19 J.J. Thomson Ave, CB3 0HE Cambridge, UK}\\
$^{16}${University of Cambridge, Kavli Institute for Cosmology, CB3 0HE Cambridge, UK}\\
$^{17}${Department of Astronomy, Nanjing University, Nanjing 210093, China}\\
$^{18}${NYU Abu Dhabi, P.O. Box 129188, Abu Dhabi, UAE}\\
$^{19}${Center for Cosmology and Particle Physics, New York University, Meyer Hall of Physics, 4 Washington Place, New York, NY 10003, USA}\\
$^{20}${Apache Point Observatory and New Mexico State University, P.O. Box 59, Sunspot, NM, 88349-0059, USA}\\
$^{21}${Sternberg Astronomical Institute, Moscow State University, Moscow}\\
$^{22}${Departamento de F\'{i}sica y Astronom\'{i}a, Facultad de Ciencias, Universidad de La Serena, Cisternas 1200, La Serena, Chile}\\
$^{23}${Department of Astronomy and Astrophysics, The Pennsylvania State University, University Park, PA 16802}\\
$^{24}${Institute for Gravitation and the Cosmos, The Pennsylvania State University, University Park, PA 16802}
\end{affiliations}
}

To illuminate the salient features of this class, we focus on a prototypical example, nicknamed ``Akira'' (Fig.~1). The SDSS imaging shows Akira to be an unremarkable spheroidal galaxy of moderate stellar mass ($\log~M_*/M_{\odot}=10.78$) that is interacting with a low-mass companion (nicknamed ``Tetsuo'') at a projected separation of $\approx32$ kpc (67$''$); they are not classified as members of a larger galaxy group\cite{yang07} and the properties of both galaxies are listed in Table 1. Spectral energy distribution (SED) fitting indicates that Akira is nearly dormant, with almost no detection of ongoing star formation\cite{chang15}. Resolved spectroscopy, however, reveals intriguing and complex patterns among spectral tracers of gas in Akira that point to a much more active internal state. With ionized gas emission detected across the entire galaxy, the map of H$\alpha$ EW (which measures the line flux relative to the stellar continuum; Fig. 1c) reveals a prominent and somewhat twisted bisymmetric pattern with a position angle (PA) of $\sim$46$^{\circ}$. The projected velocity gradient ranges from $v_{\rm ionized~gas}=-225$ km s$^{-1}$ to $v_{\rm ionized~gas}=200$ km s$^{-1}$ along the kinematic major axis, which is at a PA of $\sim$26$^{\circ}$ (Fig.~1h). We observe high ionized gas velocity dispersions across the galaxy with interesting internal structure and maxima that reach $\sigma_{\rm ionized~gas}\sim $200 km s$^{-1}$ (Fig.~1i) and $W_{80} \sim $500 km s$^{-1}$ (see Methods) perpendicular to the major kinematic axis. Meanwhile, stellar motions reveal a minimal gradient ($\pm 30$ km s$^{-1}$; Fig.~1f) that follows the PA of the galaxy's elliptical isophotes of $\sim$53$^{\circ}$ (contours in Fig.~1c). We also detect a spatially offset enhancement in Na D absorption (Fig.~1d) that is coincident with excess dust in our derived extinction map (see Methods). Measurements of the Na D line center trace a separate and distinct velocity gradient field across the offset absorption (Fig.~1e) that ranges from approximately $v_{\rm Na~D}=-80$ km s$^{-1}$ to $v_{\rm Na~D}=60$ km s$^{-1}$.

These observations indicate the presence of multiple gas components with different temperatures and velocity structures.  We interpret the ionized gas velocity field as resulting from a centrally driven (volume-filling) wind with a wide opening angle. The projected flux distribution of this ionized component largely follows the stellar surface brightness, suggesting that its primary ionization source is the local radiation field from evolved stars\cite{sarzi10, yan12, belfiore15}. The bisymmetric EW features represent enhanced emission due to shocks or over-densities along the wind's central axis.  A distinct and cooler gas component is indicated by the Na D absorption.  Because it is spatially confined with its own velocity structure, this cooler foreground material is likely to be within 1--2 $R_e$ of Akira.  Simulations of galaxy mergers constrained by the data (see Methods) suggest that the cool component is part of a tidal stream and is arcing towards the observer from the far West (blueshifted; Fig.~1e) before plunging back towards Akira's center (redshifted; Fig.~1e).

Previous work has noted similar objects\cite{sarzi10, kehrig12, allen15, gomes15} but has typically attributed their gaseous dynamics and unusual emission line features to accreted, rotating disks\cite{lagos15}. However, using a tight constraint on the total gravitational potential derived from the stellar kinematics, we find that the observed second velocity moments ($V_{\rm rms} \equiv\sqrt{V^2+\sigma^2}$) of the ionized gas in Akira are far too high to be consistent with motions under the influence of gravity alone (Fig. 1j; see Methods). Regardless of gas inclination or the degree of pressure support, we can rule out any kind of axisymmetric orbital distribution.  Perturbations or torques from disk ``settling'' are also very unlikely to drive discrepancies that reach as high as $\sim100$ km s$^{-1}$. We can express the dynamical inconsistency of the disk hypothesis another way. If we assume such a disk were inclined at $i = 50^\circ$ (see Methods), we estimate that 15-20\% of the disk would be moving at velocities sufficient to escape the galaxy. With similar velocity properties observed for the rest of theses galaxies, the disk interpretation also fails to explain why the bisymmetric $\Ha$ features are always in rough alignment with the major kinematic axis. If arising from internal structure in a moderately face-on disk, this structure should be randomly oriented compared to the kinematic axis, which is instead determined by the observer's viewing angle.

A relatively simple wind model with a constant radially-outward velocity of 310 km s$^{-1}$ confined to a wide-angle cone ($2\theta = 80^\circ$) reproduces several qualitative features in the data (Fig.~2; see Methods).  The model captures the overall shape of the ionized gas velocity field and associates the extended (horizontal) zones of high ionized gas velocity dispersion along the kinematic minor axis with the overlapping projection of approaching and receding surfaces of the inclined wind cone. By assigning somewhat greater wind densities to the cone center, we can explain the offsets between the projected kinematic major axis of the ionized gas and both the stellar position angle and the H$\alpha$ flux orientation. Furthermore, the bisymmetric H$\alpha$ EW features can be explained by enhanced gas over-densities or shock ionization along the central wind axis. Indeed, Fig.~3d-e demonstrates that line ratios in the $\Ha$ EW feature (black points and boxes throughout Fig.~3) tend to cluster and are consistent with those predicted by ``fast'' shock models\cite{allen08} with velocities of 200--400 km s$^{-1}$.

The wind's driving mechanism likely originates in Akira's radio active galactic nucleus (AGN), which is detected in FIRST (Faint Images of the Radio Sky at Twenty-Centimeters) data with a luminosity density of $L_{1.4~\rm GHz}= 1.6 \times 10^{21} ~ \rm W~Hz^{-1}$, and is most consistent with being a point source according to higher-resolution ($1.5''$) follow-up Jansky VLA (Very Large Array) radio observations (W.R., in prep). Since this AGN lacks obvious extended radio jets, its feedback is most likely manifest in small-scale jets ($< 1$ kpc) or uncollimated winds\cite{ostriker10,yuan14}. Despite an Eddington ratio of $\lambda=3.9\times10^{-4}$, energetics arguments show that the AGN's mechanical output ($P_{\rm mech} = 8.1 \times10^{41} \rm ~erg ~s^{-1}$) is sufficient to supply the wind's kinetic power ($\dot{E}_{\rm wind} \sim10^{39}$ erg s$^{-1}$; see Methods). Moreover, the wind can inject sufficient energy, coupled to the ambient gas through the turbulent dynamics observed (Fig.~1i and Fig. 3a-c), to balance cooling in both the ionized and cool gas ($\dot{E}_{\rm gas} \sim 10^{39}$ erg s$^{-1}$). Indeed, the amount of cool Na D gas ($M_{\rm cool~gas} \sim 10^8~ M_{\odot}$) implies a star formation rate of $SFR\sim1\times10^{-2}~M_{\odot} ~\rm yr^{-1}$, which is much higher than the estimated\cite{chang15} $SFR_{\rm Akira}=7\times 10^{-5}~M_{\odot}~\rm yr^{-1}$ that leverages well-detected WISE photometry. The picture that emerges is one in which cool gas inflow in Akira, triggered by the minor merger, has initiated a relatively low-power AGN-driven wind that is nonetheless able to heat the surrounding gas through turbulence and shocks and thereby prevent any substantial star formation.

As with Akira, the other galaxies in this class show little or no ongoing star formation, and the majority harbor similarly weak radio point sources (according to followup Jansky VLA observations) that would be classified as ``jet mode,'' ``kinetic mode,'' or ``radio mode'' AGN\cite{fabian12, heckman14}.  With similar levels of fast-moving ionized gas oriented along enhanced ionized emission, we conclude that AGN-driven winds are present in these systems as well and represent an important heating source.  Because the full spatial extent of these winds may exceed the field-of-view of our observations, a lower limit of $\sim$10$^7$ yr for the timescale of this phenomenon is given by the radial extent divided by the typical wind velocity.  Assuming all quiescent galaxies experience these AGN-driven winds, the $\sim$5\% occurrence rate (averaged over the full mass range) implies an episodic behavior that leads us to name these objects ``red geysers.''  Present primarily below $M_* \ls 10^{11}~M_{\odot}$, these galaxies lie in isolated halos with moderate masses\cite{yang07} ($M_{\rm halo} \sim 10^{12}~M_ {\odot}$) and exhibit no signs of major interactions.  Their implied trigger rate (at most, a few episodes per Gyr) may be related to minor mergers (approximately one per Gyr\cite{hopkins10}) as well as central accretion of ambient hot gas from stellar mass loss\cite{ciotti97}. These red geysers may represent how typical quiescent galaxies maintain their quiescence.

\noindent
\textbf{\textsf{\footnotesize Received 12 October 2015; accepted 30 March 2016.}}

\begin{addendum}
\item[Acknowledgements] We are grateful to Yu-Yen Chang for checks on the SED fitting and implied $SFR$.  We thank Stephanie Juneau, Jeffrey Newman, Hai Fu, Kristina Nyland, and S.F. S\'{a}nchez for discussions and comments.  This work was supported by World Premier International Research Center Initiative (WPI Initiative), MEXT, Japan, and JSPS KAKENHI Grant Number 15K17603. AW acknowledges support of a Leverhulme Trust Early Career Fellowship. S.P. acknowledges support from the Japan Society for the Promotion of Science (JSPS long-term invitation fellowship). M.C. acknowledges support from a Royal Society University Research Fellowship. W. R. is supported by the CUniverse Grant (CUAASC) from Chulalongkorn University.

Funding for the Sloan Digital Sky Survey IV has been provided by the Alfred P. Sloan Foundation, the U.S. Department of Energy Office of Science, and the Participating Institutions. SDSS- IV acknowledges support and resources from the Center for High-Performance Computing at the University of Utah. The SDSS web site is www.sdss.org.

SDSS-IV is managed by the Astrophysical Research Consortium for the Participating Institutions of the SDSS Collaboration including the Brazilian Participation Group, the Carnegie Institution for Science, Carnegie Mellon University, the Chilean Participation Group, the French Participation Group, Harvard-Smithsonian Center for Astrophysics, Instituto de Astrof\'{i}sica de Canarias, The Johns Hopkins University, Kavli Institute for the Physics and Mathematics of the Universe (IPMU) / University of Tokyo, Lawrence Berkeley National Laboratory, Leibniz Institut f\"{u}r  Astrophysik Potsdam (AIP), Max-Planck-Institut f\"{u}r  Astronomie (MPIA Heidelberg), Max-Planck-Institut f\"{u}r Astrophysik (MPA Garching), Max-Planck-Institut f\"{u}r  Extraterrestrische Physik (MPE), National Astronomical Observatory of China, New Mexico State University, New York University, University of Notre Dame, Observat\'{o}rio Nacional / MCTI, The Ohio State University, Pennsylvania State University, Shanghai Astronomical Observatory, United Kingdom Participation Group, Universidad Nacional Aut\'{o}noma de M\'{e}xico, University of Arizona, University of Colorado Boulder, University of Oxford, University of Portsmouth, University of Utah, University of Virginia, University of Washington, University of Wisconsin, Vanderbilt University, and Yale University.

\item[Author Contributions]
E.C. and K.B. discovered the described sources, interpreted the observations, built the wind model, and wrote the manuscript. M.C. constructed dynamical models. S.P. carried out numerical merger simulations to model the data. W.R. obtained and reduced the JVLA data. K.W. fit disk models.  K.B., R.Y., M.B., N.D., D. R. L., D. A. W., K.Z., A.W., K.L.M., and D.T. contributed to the design and execution of the survey. F.B. provided initial velocity and line-ratio maps. B.V. provided the modeled extinction map. Y.C. and K.R. contributed to the Na D interpretation.  All authors contributed to the interpretation of the observations and the writing of the paper.

\item[Author Information]
Reprints and permissions information is available at www.nature.com/reprints. The authors declare no competing financial interests. Correspondence and requests for materials should be addressed to E.C. (ec2250@gmail.com).

\end {addendum}

\begin{table*}\small
\begin{center}
\begin{tabular}{*{10}{c}}
\hline
\hline
MaNGA Name & MaNGA-ID & RA & DEC & $z^{\rm a}$ &  $\log~M_*^{\rm b}$ & $NUV-r^{\rm c}$ & $\log~SFR^{\rm d}$ & $R_{\rm e}^{\rm e}$ &  $\log~M_{\rm h}^{\rm f}$ \\ 
 & & (J2000.0 deg) & (J2000.0 deg) &    &   ($M_{\odot}$) & & ($M_{\odot}$ yr$^{-1}$) & (kpc) &  ($M_{\odot}$) \\
\hline
Akira (host) & 1-217022 & 136.08961 & 41.48174 & 0.0244671 & 10.78 & 6.1 &  -4.17 & 3.88 & 12.0 \\
Tetsuo (companion) &1-217015  & 136.11416 & 41.48621 & 0.0244647  & 9.18 &  3.0  & -0.94 & 1.73 & 12.0 \\
\hline
\end{tabular}
\caption{{\bf Galaxy properties} \newline 
{\small 
$^{\rm a}$ Spectroscopic redshift from NSA catalog. \newline
$^{\rm b}$ Stellar mass from MPA-JHU DR7 data release.\newline
$^{\rm c}$ Rest-frame $NUV-r$ color from NSA catalog.\newline
$^{\rm d}$ Star formation rate from SED fitting of SDSS optical and WISE infrared photometry\cite{chang15}; the AGN contribution to the SED is negligible.\newline
$^{\rm e}$ Effective radius from NSA catalog.\newline
$^{\rm f}$ Halo mass from a public group catalog\cite{yang07}.}}
\label{tab:geysers}
\end{center}
\end{table*}

\renewcommand{\figurename}{{\bf Figure}}

 \begin{figure*}[h!]
\centering
\includegraphics[width=183mm]{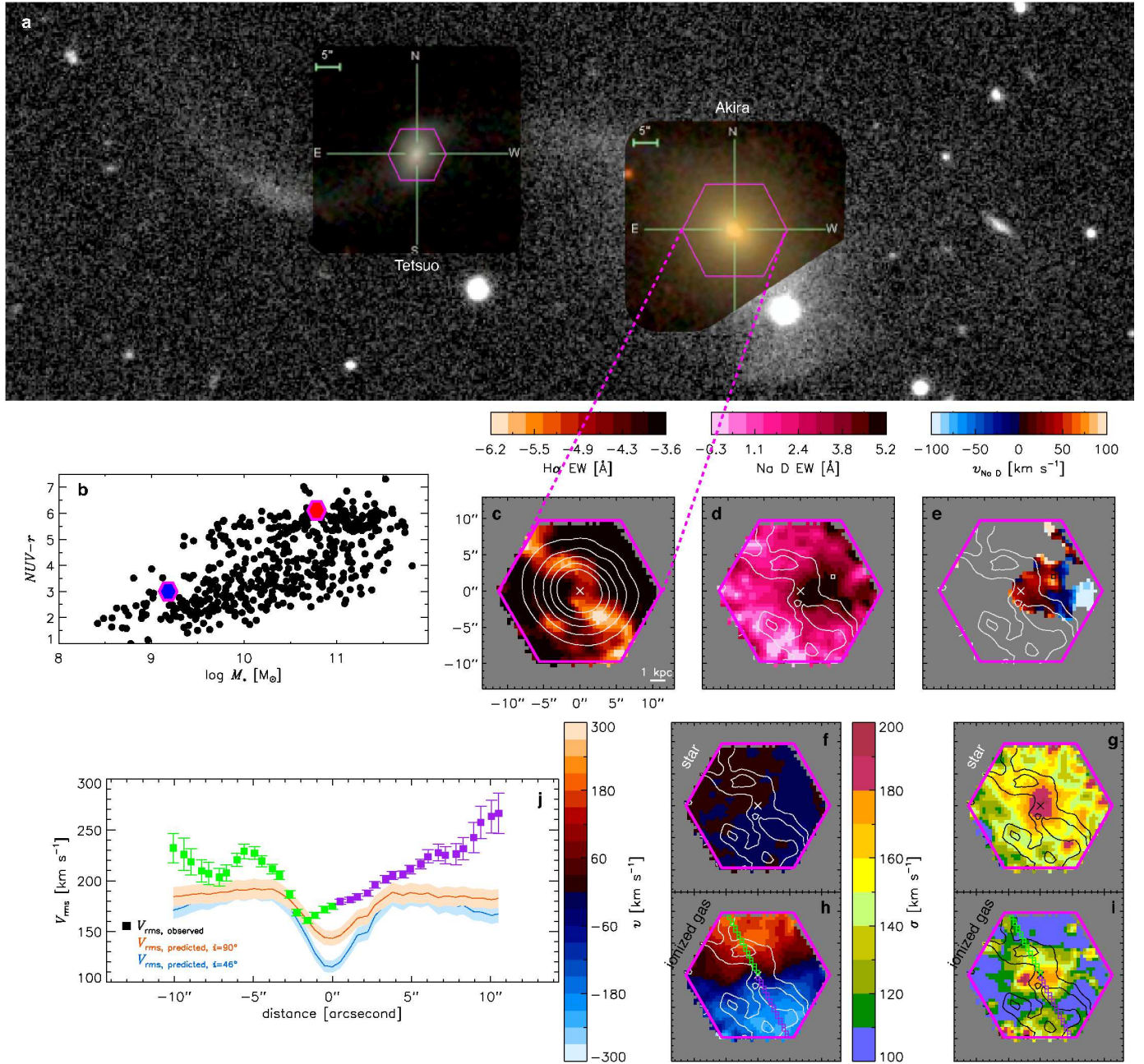}
\caption{ {\bf Akira: the prototypical red geyser.} {\bf a,} The SDSS $gri$ color images of Akira (West) and Tetsuo (East) embedded in a larger SDSS $r$ image, with the MaNGA footprint in pink. {\bf b,} The rest-frame $NUV-r$ vs. log $M_*$ diagram of the adopted MaNGA sample, with Akira and Tetsuo highlighted. {\bf c,} The $\Ha$ EW, with contours tracing the stellar continuum. {\bf d,} The Na D EW. {\bf e,} The Na D velocity. {\bf f,} The stellar velocity. {\bf g,} The stellar velocity dispersion. {\bf h,} The ionized gas velocity. {\bf i,} The ionized gas velocity dispersion. The $\Ha$ EW contours are overplotted on panels {\bf d-i}. {\bf j,} The observed $V_{\rm rms}$ from the highlighted spaxels exceeds the $V_{\rm rms}$ predicted from the gravitational potential, ruling out disk-like rotation. The error bars on the observed  $V_{\rm rms}$ represent the 1$\sigma$ measurement errors while the shaded regions around the predicted $V_{\rm rms}$ represent a conservative estimate of the systematic uncertainties.}
\label{fig:splash}
\end{figure*}

\begin{figure*}[h!]
\centering
\includegraphics[width=120mm]{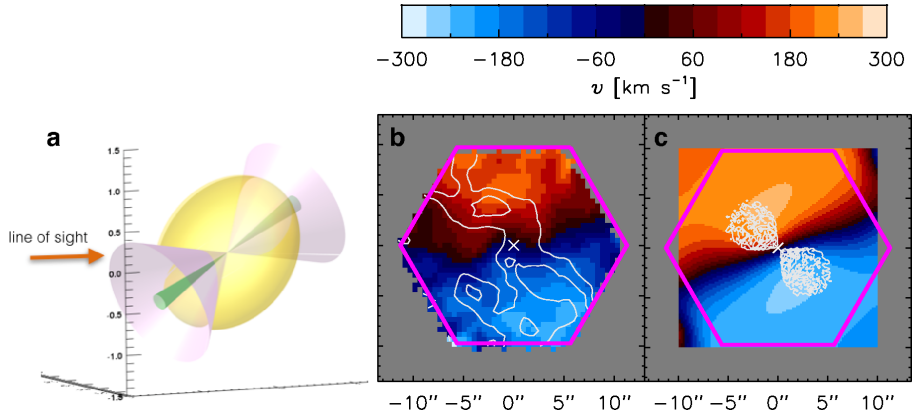}
\caption{{\bf Wind model.} {\bf a,} A schematic diagram of the galaxy (gold) and the wind bicone (purple; $2\theta = 80^\circ$) with the central $\pm$10$^\circ$ of the bicone highlighted in green. {\bf b,} Akira's $v_{\rm ionized~gas}$ map, overplotted with its $\Ha$ EW contours. {\bf c,} The projected velocity field derived from the wind model, with the white contours outlining the central axis of the wind. }
\label{fig:wind}
\end{figure*}

\begin{figure*}[h!]
\centering
\includegraphics[width=120mm]{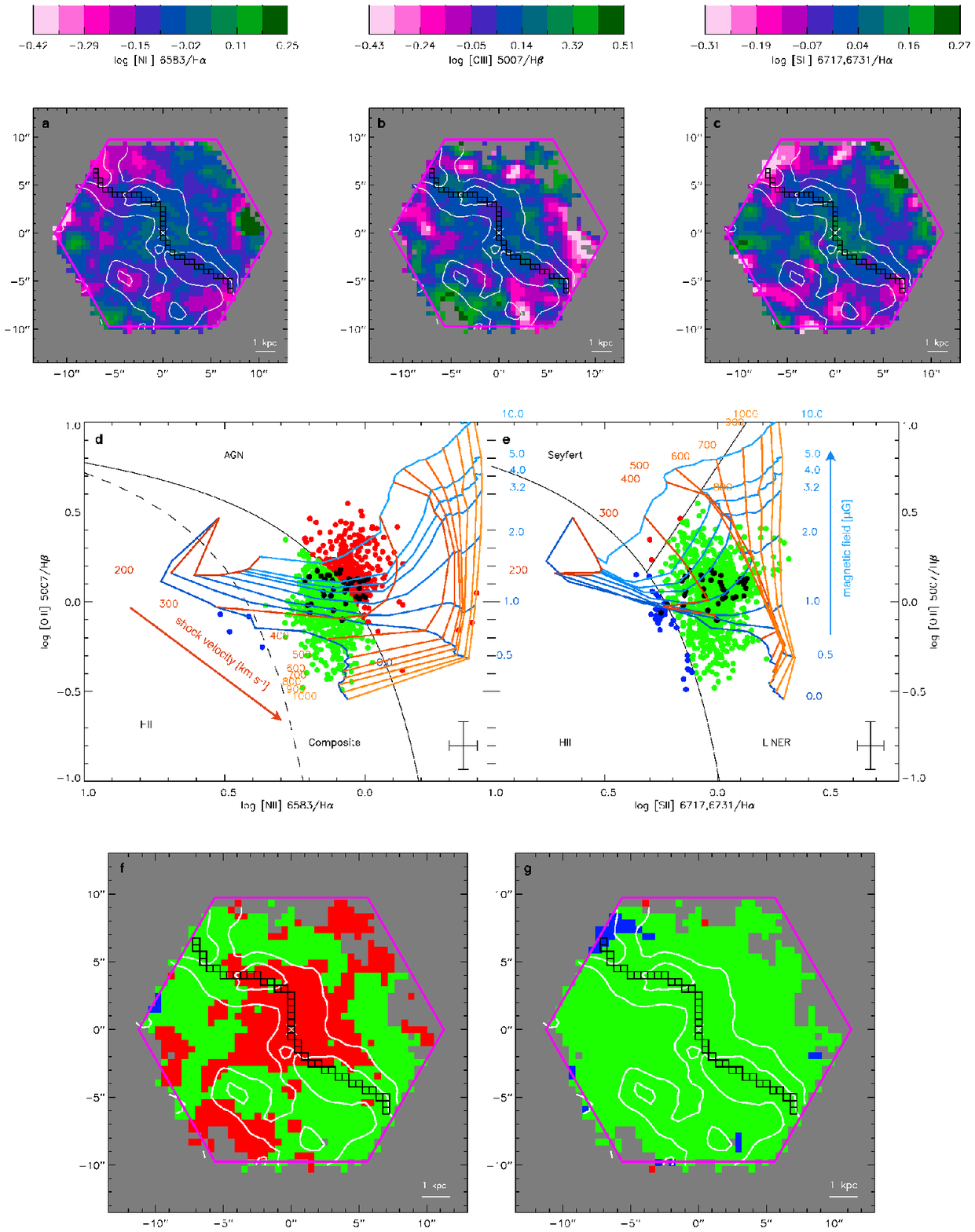}
\caption{{\bf Diagnostic line ratio maps of Akira.} {\bf a-c,} The log $\NII$ 6583/$\Ha$, log $\OIII$ 5007/H$\beta$, and log $\SII$ 6717,6731/$\Ha$ line ratio maps, with contours tracing the $\Ha$ EW pattern. {\bf d-e,} The $\NII$ 6583 and $\SII$ 6717,6731 BPT diagrams; the error bars represent the 1$\sigma$ measurement errors propagated to the log line ratios. Overplotted are shock models\cite{allen08}, and the black points correspond to the spaxels highlighted by black boxes in the other panels. {\bf f-g,} The resolved $\NII$ 6583 and $\SII$ 6717,6731 BPT maps, i.e., each spaxel is colored by its location on their respective BPT diagram.}
\label{fig:bpt} 
\end{figure*}

\clearpage

\section*{Methods}  

\subsection{Observations.} 
The data used in this work comes from the ongoing MaNGA survey\cite{bundy15,drory15,law15} using the SDSS 2.5-m telescope\cite{gunn06}. One of three programs comprising SDSS-IV, MaNGA is obtaining spatially resolved spectroscopy for 10,000 nearby galaxies with $\log~M_*/M_{\odot}\gs9$ and a median redshift of $z\approx0.04$. The $r$-band signal-to-noise ratio (S/N) in the galaxy outskirts is 4-8 $\angstrom^{-1}$, and the wavelength coverage is 3600-10,300 $\angstrom$. The effective spatial resolution is 2.4$''$ (full width at half max; FWHM) with an instrumental spectral resolution of $\sigma \sim 60$ km s$^{-1}$. The sample and data products used here were drawn from the internal MaNGA Product Launch-3 (MPL-3), which includes $\approx700$ galaxies observed before April 2015 and will be publicly available in the thirteenth SDSS data release. \newline
\indent Ancillary data are from the NASA-Sloan Atlas (NSA\cite{nsa}), MPA-JHU DR7 data release\cite{mpa}, and other recent works\cite{yang07,chang15}. We assume a flat cosmological model with $H_{0} = 70$ km s$^{-1}$ Mpc$^{-1}$, $\Omega_{m} = 0.30$, and  $\Omega_{\Lambda} =0.70$, and all magnitudes are given in the AB magnitude system\cite{oke83}. \newline
\indent The Data Analysis Pipeline (DAP), which uses {\tt pPXF}\cite{cappellari04} and the MIUSCAT stellar library\cite{vazdekis12}, fits the stellar continuum in each spaxel and produces estimates of the stellar kinematics. Flux and EW measurements were measured through simple flux-summing\cite{yan06} after we subtract the stellar continuum. We only show flux and EW measurements with S/N$>3~\angstrom^{-1}$ in the wavelength range around a given line. Ionized gas kinematics, i.e., $v_{\rm ionized~gas}$ and $\sigma_{\rm ionized~gas}$, were estimated by fitting a single Gaussian to the H$\alpha$ emission line.

\subsection{$\mathbf W_{\rm \bf 80}$.}
$W_{\rm 80}$ is a non-parametric measure of line widths; it is defined to contain 80\% of the emission-line flux\cite{harrison14}.

\subsection{Na D measurements.} 
Using a spectral fitting code\cite{fritz11, fritz14}, we present the dust extinction map of Akira in Extended Data Figure 1. The superimposed Na D contours (from Fig. 1d) overlap with enhanced extinction (darker spaxels),  supporting the association of the offset Na D absorption with cool foreground material. \newline
\indent To measure the line-of-sight (LOS) velocity of this Na D-absorbing material, which we defined as spaxels with Na D EW > 3.5 $\angstrom$, we first subtract the stellar continuum fit determined for Akira by the DAP. Extended Data Figure 2a-c shows the stellar continuum fits (red) around the Na D doublet (the two black vertical lines mark the expected locations of the Na D doublet) for three spectra. Extended Data Figure 2a shows data from a recent work\cite{chen10}, Extended Data Figure 2b shows the central spaxel of Akira, marked by the "x" in the Fig.~1d-e and Extended Data Figure~1, and Extended Data Figure 2c shows a spaxel to the Northwest of Akira, marked by the single box in the upper right of Fig.~1d-e and Extended Data Figure~1. We then examine the residual absorption as a function of wavelength, as shown in Extended Data Figure 2d-f. \newline
\indent Focusing on Extended Data Figure 2d-f, we determine the line centroids in Akira by first defining a reference Na D profile for typical, cold interstellar medium gas at rest. We use the stacked, continuum-subtracted spectrum of Na D from a large set of highly inclined disk galaxies for this reference\cite{chen10}, which is shown in the left panel. We define an at-rest line centroid for cool Na D gas by averaging the wavelengths in this profile, each weighted by the amplitude of the residual absorption at that wavelength (weighting is performed within the green region). The resulting centroid is marked by the dotted grey vertical line, which is repeated in the two right panels for reference. In the same way, we determine line centroids for the observed residual profiles across the Na D-absorbing material in Akira, which is marked by the blue vertical lines in the two right panels. We then calculate the velocity difference between the reference Na D centroid and the observed Na D centroid in these spaxels of Akira; this velocity difference is shown in the upper left in Extended Data Figure 2e-f.

\subsection{Merger simulations.}
We modeled the interaction between Akira and Tetsuo using the GADGET-2\cite{springel05} code and the methodology described in a recent work\cite{peirani10}. These simulations are constrained by the available data and contain more than 4 million particles that account for stars, dark matter, and gas (we only consider gas in Tetsuo). These simulations also include cooling, star formation, and supernova feedback, but not AGN feedback nor the proposed wind. The initial total mass merger ratio is $\sim$1:10, but because Tetsuo loses mass during the interaction, this ratio falls to $\sim$1:20 at the time most closely matching the observations (the observed stellar mass merger ratio is 1:40).  According to the best-matching viewing angle for this prograde encounter, Tetsuo starts in the foreground to the lower-right of Akira and begins arcing over the top and away from the observer (see Extended Data Figure 3a-d).  After a glancing blow with Akira, a tidal bridge is generated that loops back and passes through Akira to form the shell structure seen to the lower-right (Extended Data Figure 3d).  This snapshot at t=0.56 Gyr best matches the SDSS $r$ image (Extended Data Figure 3f), and it indicates that Tetsuo is behind Akira. Extended Data Figure 3e shows a composite stars+gas representation at this snapshot; it indicates that a stream of cool gas from Tetsuo has followed the stellar bridge that is behind Akira, penetrated close to Akira's center, emerged in front of Akira on its lower-right side, and approaches the observer. \newline
\indent The shape of the tidal bridge and shell to the south-west in the SDSS image (Extended Data Figure 3f) provides the most significant constraints on the simulation and its viewing angle.  An important cross-check is that the orientation of Tetsuo's stellar and ionized gas velocity fields (also observed by MaNGA) are reproduced as well. The geometry and velocity scale of the cool gas is similar to the observed Na D component (Fig. 1d-e), but there are differences with the observations.  Portions of the observed Na D gas appear to be falling back into Akira (redshift; Fig. 1e), but these are not seen in the simulation until a later time step. The observed cool gas orientation is also more horizontal while the simulation predicts the gas stream stretches further (Extended Data Figure 3e). But we emphasize that we only detect cool gas in absorption where there are background stars from host galaxy, whereas the simulation allows us to see the full extent of the cool gas. Differences between the simulations and observations may also arise from inaccuracies in the initialization of the merger simulation (mass ratios, gas mass fractions, angular momentum alignment, etc.), limitations in the hydrodynamic gas treatment, or missing components in the simulation such as Akira's gas supply and the proposed AGN-driven wind.

\subsection{Dynamical modeling evidence against the presence of disks.}
Jeans Anisotropic Modeling (JAM\cite{cappellari08}), which uses the Multi-Gaussian Expansion (MGE\cite{emsellem94,cappellari02}) parametrization for mass and light distributions, was performed on Akira and other red geysers to model their stellar kinematics and gravitational potential. The JAM model derives a 3D stellar density by de-projecting the observed SDSS $r$-band photometry using an MGE fit. The modeled potential includes an NFW\cite{nfw96} dark matter halo. The JAM model has four free parameters: the inclination $i$, anisotropy $\beta_z$, stellar $M/L$, and halo mass. These are optimized by fitting the model prediction for the second velocity moments, $V_{\rm rms}\equiv\sqrt{V^2+\sigma^2}$, to the observed MaNGA stellar kinematics. Through a number of systematics tests, we find that the best-fit stellar inclination is $i=41^\circ$, with an upper limit of $i= 50^\circ$. Although there is some covariance between the model parameters, the resulting {\em total} mass profile, is extremely robust\cite{li16}.

With the total gravitational potential defined from the stellar JAM modeling above, we can predict projected second velocity moment ($V_{\rm rms}$) maps of gas under the assumption of axisymmetric orbital distributions. We treat the $\Ha$-emitting gas clouds as a ``tracer'' population of the underlying potential. Its flux distribution is modeled by a separate MGE (distinct from the stellar component) enabling de-projection of the observed $\Ha$ surface brightness. The Jeans equations are then solved for this tracer, within the fixed potential, to predict the $V_{\rm rms}$ allowed by the given mass distribution. We emphasize that the second moments are independent of the degree of circular motion versus ``random'' motion in the hypothesized disk. The analysis does not account for non-gravitational drivers of turbulent pressure, such as from the AGN-driven wind we propose. In Extended Data Figure 4 (see also Fig. 1j) we show results for gaseous inclinations of $i=46^\circ$ (the minimum allowed by the $b/a=0.7$ from GALFIT fits of the $\Ha$ flux, corresponding to an intrinsic axis ratio $q=0.12$; see below) and the most extreme case of $i=90^\circ$ (an edge-on axisymmetric density). In either case, the allowed $V_{\rm rms}$ is far below the observed $V_{\rm rms}$.

With discrepancies as high as $\sim100$ km s$^{-1}$, torques of the same order as the gravitational potential itself would be required to explain the data, making a ``disturbed'' disk a highly unlikely explanation. It is possible to imagine a very chaotic accretion scenario where the JAM assumptions of axisymmetry and stability completely break down, although in this case an ordered velocity gradient of the kind observed seems unlikely.  Such a scenario would also struggle to explain how the high dispersions are generated and why enhanced $\Ha$ flux is observed along the gradient in the gradient field. Similarly, because line widths of $W_{80} \sim $500 km s$^{-1}$ could not be sustained by accreting tidal streams or caused by tidal torques, multiple overlapping gas streams would have to conspire to produce the widespread high velocity dispersion observed (Fig.~1i) while maintaining an ordered velocity gradient pattern. A similar set of coincidences would be required for each galaxy in the rest of the red geyser sample.

Not surprisingly, tilted-disk models\cite{andersen13} that fit the ionized velocity field alone do a poor job for the red geyser sample.  Characterizing the goodness-of-fit by an error-weighted average residual, the majority of red geysers exhibit residuals that place them among the worst 5\% of fitted MaNGA galaxies with ``disk-like'' kinematics.  Here, disk-like refers to galaxies with reasonable agreement between stellar and gaseous systemic velocities, dynamical centers, position angles, and inclinations. 

Finally, we use the dynamically-constrained potential to estimate a local escape velocity and compare this to the inferred velocity distribution of a putative disk.  Several assumptions are required, but the results are informative.  We obtain a rough estimate of escape velocity, $v_{\rm esc} \sim 400 \pm 50$ km s$^{-1}$, by integrating the potential from a projected radius of $7''$ (3.4 kpc or just under 1 $R_{\rm e}$) to 4 $R_{\rm e}$ (16 kpc) and assuming a gentle decline in the circular velocity at large radius. We then use GALFIT\cite{peng02} to model the observed $\Ha$ flux surface brightness, finding a consistent projected axis ratio of $b/a = 0.7 \pm 0.02$, regardless of the assumed model profile (exponential, de Vaucouleurs', or free S\'ersic) and despite significant structure in the residuals (of order $\sim10-15\%$).  Hypothesizing a disk with an intrinsic axis ratio, $q = 0.4$, roughly twice as ``fat'' as typical disks\cite{bershady11}, we estimate an inclination of $i = 50^\circ$.  This is also the upper limit of inclinations allowed for the stellar kinematics, and precession should align accreted material with the stellar distribution in roughly a few dynamical times\cite{tohline82} (unless there is a source of incoming misaligned gas\cite{vandevoort15}).  We de-project the observed mean velocities using this inclination and consider the distribution of velocities about this mean.  Roughly 15-20\% of the gas, i.e., with velocities greater than 1$\sigma$ from the mean, would exceed the escape velocity under these assumptions.

\subsection{Wind model.}
We construct a simple wind model that reproduces many qualitative features of the MaNGA observations.  In this model, the wind assumes a wide-angle biconical form centered on the galaxy nucleus.  Within the bicone, the wind has a constant amplitude, radially-outward velocity\cite{bouche12}.  We assume that warm gas clouds entrained by the wind trace this velocity structure and emit flux in strong emission lines primarily in response to the local ionization field supplied by the stars\cite{sarzi10, yan12, belfiore15}.  The projected wind velocity field to first order is therefore a convolution of the wind geometry with the galaxy's 3D luminosity profile.

To realize the model, we populate a randomized 3D cartesian grid of points with the galaxy at the center and assign each point a weight equal the value of an axisymmetric Hernquist density profile sampled at that point\cite{dehnen94}.  This density profile is fixed to reproduce the imaging and JAM constrains on the stellar component, namely an intrinsic (3D) axis ratio of 0.4, an inclination of 41$^\circ$, a projected major-axis effective radius of $R_e \approx 7''$, and an on-sky PA of 53$^\circ$.  For a given wind opening angle and inclination, we weight the projected line-of-sight component of the wind velocity at each point inside the bicone by its Hernquist profile value.  Projected quantities are smoothed to the spatial resolution of the MaNGA data (2.4$''$, FWHM).  To model a potential enhancement of gas densities or shocks along the central axis of the bicone, we implement a second set of weights defined with respect to the bicone that decrease exponentially (with a variable characteristic angle) as a function of the angular distance from the bicone's axis.

By experimenting with different choices for the wind's opening angle, inclination, length, intrinsic velocity, (and central weighting, if desired), we explored possible wind model solutions.  Most have opening angles of $2\theta \sim 80^\circ$ and steep inclinations ($\sim$70$^\circ$) toward the line-of-sight.  One example is shown in Fig.~2. This wind model has an opening angle of $2\theta = 80^\circ$, an inclination of 75$^\circ$, PA = 55$^\circ$, and a length of $2 R_e$. We have assumed a constant radially outward velocity within the wind of $v_{\rm wind} = 310$ km s$^{-1}$.  We associate the observed, bisymmetric regions of enhanced H$\alpha$ (white contours on the observed velocity field; Fig.~2b) with the wind's central axis.  The projection of this $\pm$10$^\circ$ region is overplotted with white contours on the modeled velocity field (Fig.~2c).  The wind density is assumed to decline as an exponential function of the angular distance with a characteristic angle of $\alpha = 10^\circ$.

\subsection{Shock models.}
Shock models\cite{allen08} with twice the solar atomic abundances, shock velocities of 200-400 km s$^{-1}$, magnetic fields of 0.5-10 $\mu$G, and preshock densities of unity, were used in Fig.~3.

\subsection{Inferring the presence of an AGN in Akira.}
The presence of a central radio source and the absence of star formation in Akira imply the presence of an AGN. Quantitatively, we can confirm the presence of an AGN by comparing the expected $SFR$ inferred from the radio luminosity of Akira to the estimated $SFR$ from SED fitting of SDSS and WISE photometry\cite{chang15}. We first calculate the radio luminosity density of Akira using $L_{1.4~\rm GHz} = 4\pi d_{\rm L}^{2}  F_{1.4~\rm GHz}$, where $F_{1.4~\rm GHz}$ is the integrated flux density (of 1.2 mJy) from FIRST\cite{becker95}, and $d_{\rm L}$ is the luminosity distance. This calculation yields $L_{1.4~\rm GHz}= 1.6 \times 10^{21} ~  \rm W~Hz^{-1}$. Using the radio star formation rate calibration\cite{kennicutt12}, we infer $SFR = 1~M_{\odot}~\rm yr^{-1}$. This level of star formation in Akira is ruled out at more than $97.5\%$ confidence\cite{chang15}, indicating that the most likely source of this radio emission is an AGN.

\subsection{Eddington ratio and AGN power.}
To calculate the Eddington ratio ($\lambda$) of Akira, we use the Eddington-scaled accretion rate\cite{best12}, which is more applicable to radio-detected AGN: $\lambda=(L_{\rm rad} + L_{\rm mech})/L_{\rm Edd}$, where $L_{\rm rad}$ is the bolometric radiative luminosity, $L_{\rm mech}$ is the jet mechanical luminosity, and $L_{\rm Edd}$ is the Eddington limit. To calculate $L_{\rm rad}$, we converted the $\OIII$ 5007 flux from the central $2''$ ($\approx1$ kpc) radius aperture of Akira, $F_{\rm \OIII}$, to a luminosity: $L_{\rm \OIII} = 4\pi d_{\rm L}^{2}  F_{\rm \OIII}=1.7 \times 10^{39}\rm ~erg~s^{-1}$. Even though the central $\OIII$ 5007 flux is probably not entirely due to AGN photoionization (evolved stars and shocks probably contribute), for this order-of-magnitude calculation we will make the simplifying assumption that it does. Using the relation\cite{heckman04} $L_{\rm rad} = 3500L_{\rm \OIII}$, we obtain $L_{\rm rad} =  5.9 \times 10^{42} \rm ~erg~s^{-1}$. 

Acknowledging that Akira is in a lower mass and energy output regime than those in which expanding X-ray bubbles have been observed, we nonetheless applied the following relation\cite{cavagnolo10} to calculate the jet mechanical luminosity: $L_{\rm mech}=7.3\times10^{36}(L_{\rm 1.4 ~ GHz}/10^{24} ~ \rm W~Hz^{-1})^{0.70} \rm~W$, which results in $L_{\rm mech} = 8.1\times10^{34}~W = 8.1 \times10^{41} \rm ~erg ~s^{-1}$.

Finally, to calculate $L_{\rm Edd}$, we first estimate the black hole mass, $M_{\rm BH}$, using the relation\cite{mcconnell13} $\log(M_{\rm BH}/M_{\odot})=8.32+5.64\log[\sigma_{\rm star}/(200~\rm km~s^{-1})]$, with $\sigma_{\rm star}= 185.5~\rm km~s^{-1}$ from the central $2''$ radius aperture, yielding $\log(M_{\rm BH}/M_{\odot})=8.1$. We calculate the classical Eddington limit with $L_{\rm Edd}=3.3\times10^{4}M_{\rm BH}= 4.5\times10^{12}~L_{\odot} = 1.7\times10^{46}~\rm erg~s^{-1}$.

Inserting these numbers into $\lambda=(L_{\rm rad} + L_{\rm mech})/L_{\rm Edd}$ yields $\lambda=3.9\times10^{-4}$, suggesting that the accretion onto this black hole is at a low rate and/or radiatively inefficient; these types of AGN have been termed low-energy, kinetic mode, jet mode, or radio mode AGN\cite{fabian12, best12, heckman14}. 

\subsection{Ionized gas energetics.} 
Assuming warm ionized gas clouds with a temperature of $10^4$ K and using the observed \SII~ratio, we estimate\cite{osterbrock89} an electron density, $n_e$, of 100 cm$^{-3}$.  With this value of $n_e$, we estimate\cite{genzel11} the lower limits on the ionized gas mass from the H$\alpha$ line flux, $M_{\rm warm,~H\alpha} \sim 6 \times 10^5$ $\msun$. We can derive similar estimates\cite{carniani15} based on the H$\beta$ and \OIII~flux, obtaining $M_{\rm warm,~H\beta} \sim 4 \times 10^5~\msun$ and $M_{\rm warm,~\OIII} \sim 2 \times 10^4$ $\msun$.  We adopt an approximate $M_{\rm warm} \sim 10^5$ $\msun$.

To approximate the energy associated with a wind driving the observed velocities in the ionized gas, we adopt the kinetic energy\cite{harrison14}, $E_{\rm wind} \sim \frac{1}{2}M_{\rm warm} v_{\rm wind}^2$, with $v_{\rm wind} = 300$ km s$^{-1}$. To estimate the wind power, we divide $E_{\rm wind}$ by the characteristic wind timescale of 10$^7$ yr, derived by dividing the Akira's optical radius (the observed extent of the wind) by $v_{\rm wind}$.  We obtain $\dot{E}_{\rm wind} \sim 10^{39}$ erg s$^{-1}$. Because the ionized gas mass is likely a lower limit, $\dot{E}_{\rm wind}$ is likely an underestimate. The gas cooling rate is estimated using a method from the literature\cite{sutherland93}.  

\subsection{Star formation in the Na D cool gas.}  
To estimate the expected star formation rate associated with the cool Na D gas, we first estimate\cite{bohlin78} the total hydrogen column density (N$_{HI} + 2N_{H_2}$) from the dust extinction presented in Extended Data Figure~1.  Integrating over the $\sim$4 kpc$^2$ region of enhanced extinction, we find a total gas mass of $M_{\rm cool} \sim 10^8$ $\msun$ or a surface mass density of $\Sigma_{\rm cool} \sim 3 \times 10^7~\msun$ kpc$^{-2}$.  To apply the Kennicutt relation\cite{kennicutt07}, we first account for fact that the Na D material is unlikely to be distributed in a thin, face-on disk.  Assuming the Kennicutt relation holds with respect to \emph{volumetric} density, we scale $\Sigma_{\rm cool}$ by the ratio of scale heights between a typical star-forming spiral ($H_{\rm Kennicut} \sim 0.6$ kpc\cite{bershady10}) and an estimate for the Na D material's scale height, $H_{\rm NaD}$.  We set $H_{\rm NaD}$ to $\sim$3 kpc, which is approximately the effective radius ($R_{\rm e}$) of Akira.  These assumptions yield $SFR\sim10^{-2}~M_{\odot} ~\rm yr^{-1}$, roughly 100 times higher than the estimate for Akira ($SFR_{\rm Akira}=7\times 10^{-5}~M_{\odot}~\rm yr^{-1}$)\cite{chang15}.

\subsection{Code availability.} 
The JAM code is available at \url{http://www-astro.physics.ox.ac.uk/~mxc/software/}.




\renewcommand{\figurename}{{\bf Extended Data Figure}}
\setcounter{figure}{0}

\begin{figure*}
\centering
\includegraphics[width=89mm]{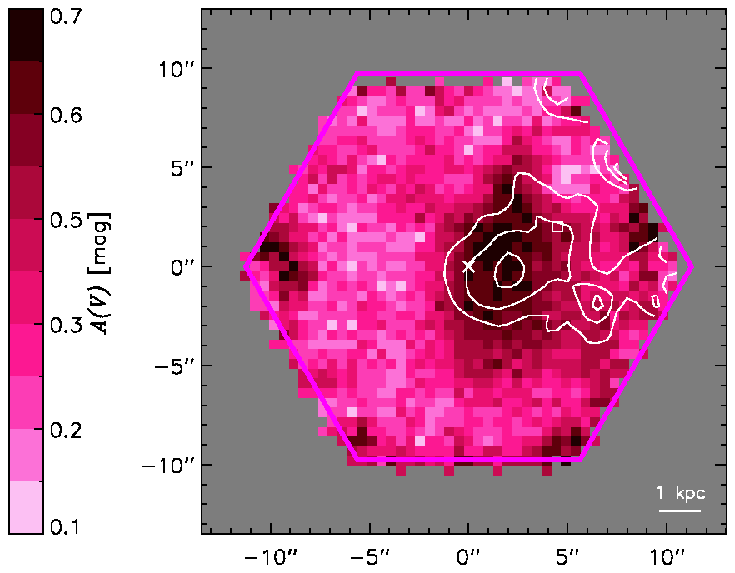}
\caption{{\bf $\mathbf{A}(\rm \bf \it V)$ map.} The estimated $A(V)$ map, with contours of Na D EW > 3.5 $\angstrom$ from Fig.~1d. The spatial overlap between regions of high extinction and the Na D EW absorption confirms that there is cool material in the foreground of Akira.}
\label{fig:av}
\end{figure*}

\begin{figure*}
\centering
\includegraphics[width=183mm]{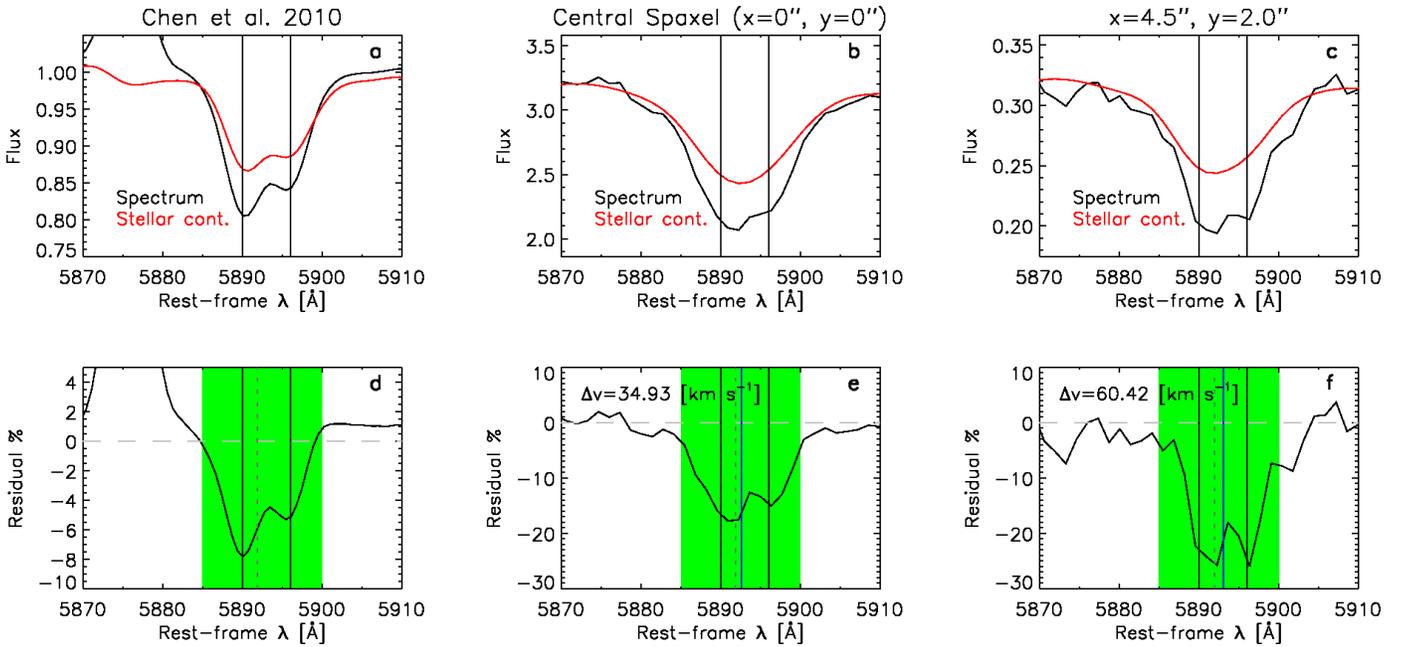}
\caption{{\bf Na D LOS measurement.} {\bf a-c,} The spectrum around the Na D doublet at $\lambda=5890, 5896~\angstrom$ and best-fit stellar continuum. The two vertical lines mark the locations of the Na D doublet. {\bf d-f,} The residual of the spectrum and stellar continuum. Considering only the wavelength range enclosed by the green region, we calculate the residual-weighted central wavelengths of these Na D doublets, which is marked by the vertical lines (the dashed grey vertical represents the reference Na D centroid while the blue vertical lines represent the observed Na D centroid from the two spaxels of Akira).}
\label{fig:nad_los}
\end{figure*}

\begin{figure*}
\centering
\includegraphics[width=183mm]{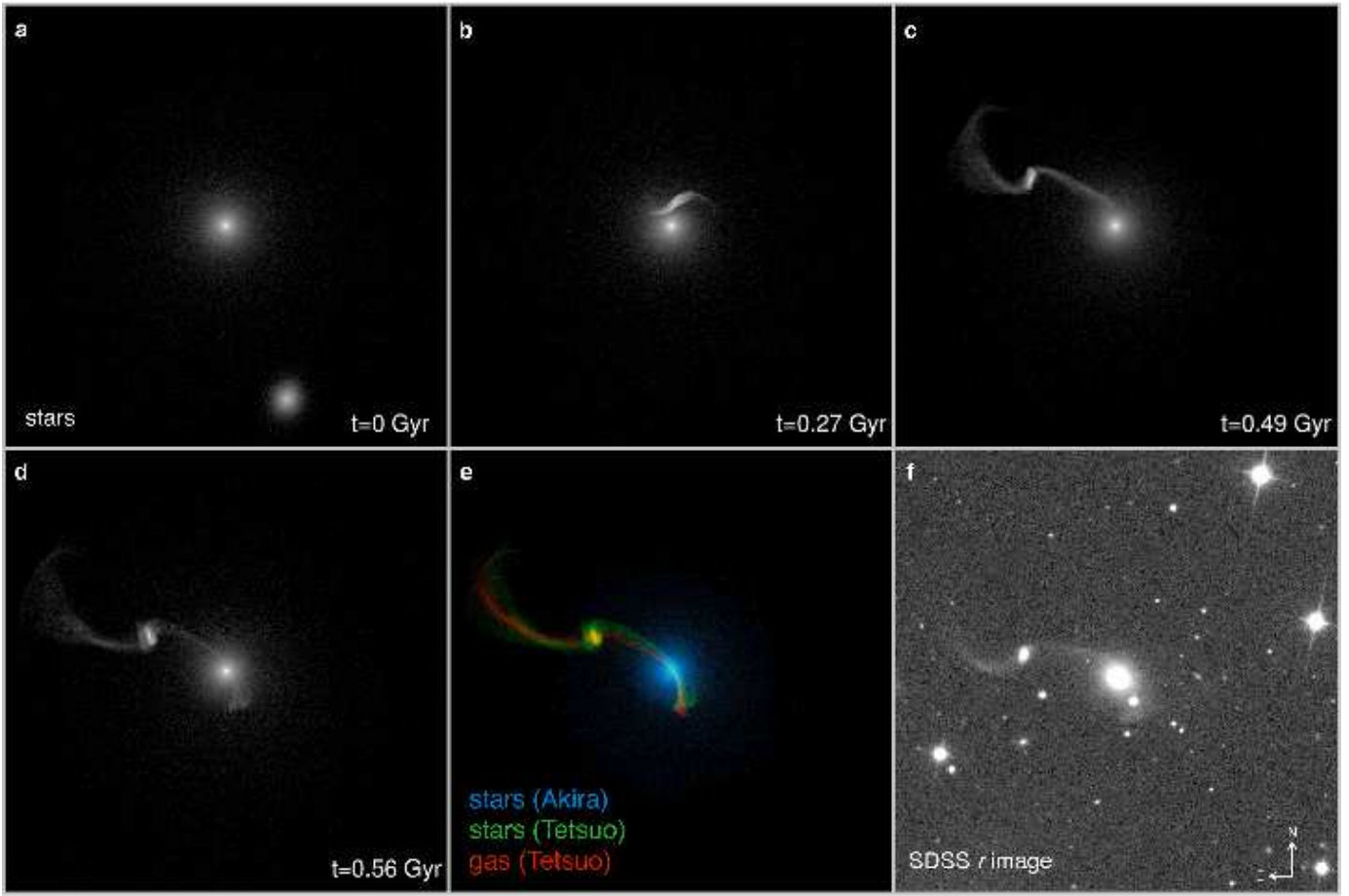}
\caption{{\bf Merger simulation.} {\bf a-d,} Evolution of the stars from t=0 Gyr to t=0.56 Gyr; each panel is $90\times90$ kpc. {\bf e,} Composite image of stars and gas at t=0.56; this panel is also $90\times90$ kpc {\bf f,} The SDSS {\it r} image of Akira and Tetsuo.}
\label{fig:merger_stars}
\end{figure*}

\begin{figure*}
\centering
\includegraphics[width=183mm]{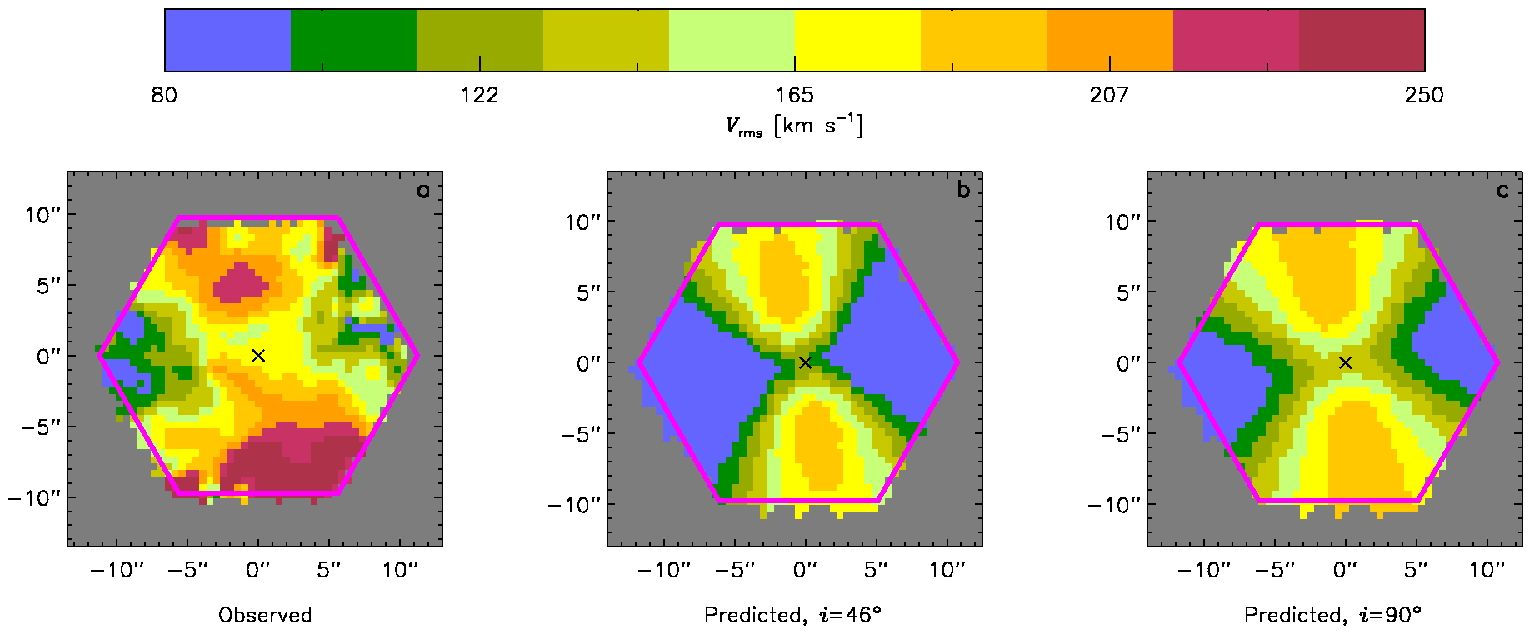}
\caption{{ $\mathbf \it V_{\rm \mathbf{rms}}$ \bf maps.} {\bf a,} Observed $V_{\rm rms}$ map. {\bf b,} Predicted $V_{\rm rms}$ map, assuming $i=46^\circ$. {\bf c,} Predicted $V_{\rm rms}$ map, assuming $i=90^\circ$.}
\label{fig:vrms}
\end{figure*}

\end{document}